\documentclass[aps,twocolumn,showpacs,floatfix]{revtex4}

\usepackage{graphicx}
\usepackage{amsmath}
\usepackage{amsfonts}
\usepackage{amssymb}

\begin{document}

\def\be{\begin{equation}}
\def\ee{\end{equation}}
\def\bea{\begin{eqnarray}}
\def\eea{\end{eqnarray}}
\def\bma{\begin{mathletters}}
\def\ema{\end{mathletters}}
\newcommand{\one}{\mbox{$1 \hspace{-1.0mm}  {\bf l}$}}
\newcommand{\eins}{\mbox{$1 \hspace{-1.0mm}  {\bf l}$}}
\def\C{\hbox{$\mit I$\kern-.7em$\mit C$}}
\newcommand{\tr}{{\rm tr}}
\newcommand{\half}{\mbox{$\textstyle \frac{1}{2}$}}
\newcommand{\shalf}{\mbox{$\textstyle \frac{1}{\sqrt{2}}$}}
\newcommand{\ket}[1]{ | \, #1  \rangle}
\newcommand{\bra}[1]{ \langle #1 \,  |}
\newcommand{\proj}[1]{\ket{#1}\bra{#1}}
\newcommand{\kb}[2]{\ket{#1}\bra{#2}}
\newcommand{\bk}[2]{\langle \, #1 | \, #2 \rangle}
\def\II{I(\{p_k\},\{\rho_k\})}
\def\ss{{\cal K}}
\tolerance = 10000

\bibliographystyle{apsrev}

\title{From Cooper pairs to Luttinger liquids with bosonic
atoms in optical lattices}

\author{B. Paredes}
\author{J. I. Cirac}

\affiliation{Max--Planck Institute for Quantum Optics,
Hans--Kopfermann Str. 1, D-85748 Garching, Germany}

\begin{abstract}

We propose a scheme to observe typical strongly correlated {\em
fermionic} phenomena with {\em bosonic} atoms in optical lattices.
For different values of the sign and strength of the scattering
lengths it is possible to reach a "superconducting" regime, where
the system exhibits atomic pairing, or a Luttinger liquid
behavior. We identify the range of parameters where these
phenomena appear, illustrate our predictions with numerical
calculations, and show how to detect the presence of pairing.

\end{abstract}

\date{\today}
\pacs{03.75.Fi, 03.67.-a, 42.50.-p, 73.43.-f } \maketitle

After the achievement of Bose--Einstein condensation with dilute
atomic gases \cite{bec}, a great deal of interest in atomic
physics has turned to the theoretical and experimental study of
cold fermionic atoms. In particular, one of the most challanging
goals nowadays is the observation of the BCS transition
\cite{bcs1}, where fermionic atoms are expected to form Cooper
pairs when they experience an attractive interaction. Although
Fermi degeneracy has already been observed in several labs
\cite{fer}, the temperatures at which Cooper pairs are formed have
not been reached (see, however, \cite{bcs2}).

In this letter we propose to use {\em bosonic} atoms trapped in
optical lattices to observe typical {\em fermionic} correlation
phenomena, such the formation of Cooper pairs or the
characteristic spin--density separation corresponding to a
Luttinger liquid. Our scheme is based on the fact that strongly
interacting bosons in a lattice can behave as weak (or even
strong) interacting fermions. Our proposal is motivated by the
recent experiment with bosons in optical lattices \cite{Bloch},
where the strong interaction regime has been achieved, as it was
theoretically predicted \cite{Jaksch}. In fact, this experiment
illustrates that our proposal can be implemented with existing
technology.

The fact that strongly interacting bosons can behave as fermions
is, of course, not a new idea. It is well known that the problem
of one dimensional hard core bosons is exactly mapped into the one
of {\em free} fermions \cite{Girardeau}, since the infinite on
site repulsion plays the role of an effective Pauli principle for
bosons. This is the case, for instance, for the Mott phase of
bosonic atoms in an optical lattice \cite{Bloch,Jaksch}, or for
the Laughlin atomic liquids in rapidly rotating traps
\cite{belen}. In an attempt to go beyond the usual free-fermionic
behavior of hard core bosons, we show that the ability of tuning
the sign and strength of atomic interactions can easily make
bosons behave as {\em interacting} fermions. To be more specific,
we will consider bosonic atoms confined in an optical lattice in
such a way that tunneling can only occur in one spatial dimension.
We will assume that some internal levels have been shifted up in
energy using off resonant microwave fields, so that only two
internal levels can be occupied. We will concentrate in the
situation in which the interaction between atoms in the same
internal level, $U$, is repulsive and stronger than the
interaction between atoms in different ones, $V$ \cite{footnotea}.
The basic mechanism behind our scheme is simple. At sufficient low
temperatures and tunneling amplitude, the strong repulsion $U$
prevents two atoms with the same internal state to be in the same
lattice site, so that atoms behave as interacting fermions. We
will show that the interaction between the effective fermions is
just the bare interaction $V$, and depending on its sign and
strength a rich spectrum of possibilities can be observed.

We consider a gas of $N$ bosonic atoms in a one-dimensional
optical lattice with $M$ wells, and will call $\nu=N/2M$ the
filling factor.  We will denote by $\sigma={\uparrow, \downarrow}$
the two relevant internal levels. The Hamiltonian describing this
situation is \cite{Jaksch}:
\begin{equation}
H=-t \sum_{<i,j>\sigma}a^{\dagger}_{i\sigma}a_{j\sigma}+
U\sum_{i,\sigma}n_{i \sigma}^2 +V\sum_{i}n_{i \uparrow}n_{i
\downarrow}. \label{hamb}
\end{equation}
Here $a^{\dagger}_ {i\sigma}$ and $a_{i\sigma}$ are bosonic
operators that create (annihilate) an atom on the $i$--th lattice
site with spin state $\sigma$, and $n_{i
\sigma}=a^\dagger_{i\sigma}a_{i\sigma}$. The tunneling amplitude,
$t$, as well as $U$ and $V$ can be easily written in terms of the
scattering lengths and lattice parameters \cite{Jaksch}.

Let us analyze first the simple limiting case $U\to \infty$, where
the atoms with the same spin are not allowed to be at the same
lattice site. The ground state and low-energy excitations of
Hamiltonian (\ref{hamb}) lie within the subspace generated by
states of the form:
\begin{equation}
|\mathbf{x},\mathbf{y} \rangle=a^{\dagger }_{x_{1} \uparrow} \ldots
a^{\dagger }_{x_{N_{\uparrow}} \uparrow}\,
a^{\dagger }_{y_{1} \downarrow}\ldots
a^{\dagger }_{y_{N_{\downarrow}} \downarrow}|0 \rangle,
\label{sub}
\end{equation}
where $N_{\uparrow}+N_{\downarrow}=N$,
$\mathbf{x}=(x_{1},\ldots,x_{N_{\uparrow}})$, and
$\mathbf{y}=(y_{1},\ldots,y_{N_{\downarrow}})$ denote,
respectively, the sites occupied by the particles with spin up and
spin down, and $x_{i} \ne x_{j}$, $y_{i} \ne y_{j}$, $\forall i
\ne j$. This projected bosonic Hilbert space is isomorphic to the
Hilbert space of fermions with spin $1/2$ in a one-dimensional
lattice. The bosonic operators can be transformed into fermionic
ones by the well known Jordan-Wigner transformation \cite{Jordan}:
\begin{equation}
a^{\dagger}_{i \sigma}=
(-1)^{\sum_{j<i} (1-2c^{\dagger}_{j\sigma}c_{j\sigma})}c^{\dagger}_{i \sigma},
\label{J-W}
\end{equation}
where the operator $c^{\dagger}_{j\sigma}$ creates an effective
fermion in the $j$--th site with spin $\sigma$. With the
transformation (\ref{J-W}) the bosonic Hamiltonian (\ref{hamb})
projected to the subspace (\ref{sub}) is exactly transformed
\cite{footnotec} into a Hubbard chain of fermions:
\begin{equation}
H_{F}=-t \sum_{<i,j>\sigma}c^{\dagger}_{i\sigma}c_{j\sigma}
+V\sum_{i}c^{\dagger}_{i\uparrow} c_{i\uparrow}
c^{\dagger}_{i\downarrow} c_{i\downarrow},
\label{hamf}
\end{equation}
where  the parameters $t$ and $V$ coincide with those appearing in
Hamiltonian (\ref{hamb}).

Given the fact that the projected Hamiltonian and (\ref{hamf}) are
the same, they have exactly the same spectrum. The bosonic
eigenstates are related to the fermionic eigenstates by the
following correspondence: $ |\Psi_{F} \rangle =
\sum_{\mathbf{x},\mathbf{y}} f_{F}(\mathbf{x},\mathbf{y})\,\, |
\mathbf{x},\mathbf{y} \rangle_{F}$ $\rightarrow$ $
\sum_{\mathbf{x},\mathbf{y}} f_{B}(\mathbf{x},\mathbf{y})\,\, |
\mathbf{x},\mathbf{y} \rangle=|\Psi_{B} \rangle$. Here, $|
\mathbf{x},\mathbf{y} \rangle_{F}= c^{\dagger }_{x_{1} \uparrow}
\ldots c^{\dagger }_{x_{N_{\uparrow}} \uparrow} c^{\dagger
}_{y_{1} \downarrow}\ldots c^{\dagger }_{y_{N_{\downarrow}}
\downarrow}|0 \rangle$, $| \mathbf{x},\mathbf{y} \rangle$ is given
by (\ref{sub}),  and $f_{F}(\mathbf{x},\mathbf{y})$,
$f_{B}(\mathbf{x},\mathbf{y})$, the fermionic and bosonic
amplitudes for the configuration $(\mathbf{x}, \mathbf{y})$, are
related by a sign factor in the form
$f_{B}(\mathbf{x},\mathbf{y})=\mathrm{sgn}(\mathbf{x}, \mathbf{y})
f_{F}(\mathbf{x},\mathbf{y})=
\prod_{i<j}\mathrm{sgn}(x_{i}-x_{j})\mathrm{sgn}(y_{i}-y_{j})\,
f_{F}(\mathbf{x},\mathbf{y})$.

The physics of the fermionic Hubbard chain has been extensively
studied (see, e.g., \cite{hub1}). Depending on the sign of $V$ and
on the ratio $V/t$ the system is known to exhibit different
phenomena. In order to determine which of these phenomena can be
observed with a real bosonic system we have to address two
questions. First, since $U$ cannot be infinite, we have to
determine the conditions on $U$ such that the above treatment
remains valid. Second, we have to analyze the effect of the
boson--fermion transformation (\ref{J-W}) on the predicted
phenomena.

Let us first estimate under which conditions our treatment will
remain valid for finite $U$. Since we are interested in the
phenomena related to the ground state (and may be low--energy
excitations) we have to compare the energy of these excitations
with the one related to leaving the projected bosonic subspace.
The first one can be determined from the known results for
fermions \cite{hub1}, whereas the second one will be of the order
of $U$. Thus, we arrive at the condition $U \ll
\mathrm{min}(t,|V|)$.

Now, let us discuss which fermionic phenomena can be observed with
the bosonic system, i.e. the effects of transformation
(\ref{J-W}). These phenomena are characterized by the nature of
the excitation spectrum and by some special behavior of the
correlation functions. In particular, for $V>0$ and $\nu \ne 1/2$
we have a Luttinger liquid \cite{Haldane} where charge and spin
excitations are independent, gapless, and with phononic dispersion
relations, leading to two different velocities. For $V>0$ and
$\nu=1/2$ there is a gap for charge excitations and the system is
a Mott insulator. For $V<0$ the system is a superconductor.
Fermions get paired, opening a gap for spin excitations. As the
strength of the attractive interaction increases, the system
evolves continuously from cooperative Cooper pairing (the BCS
regime) to independent bound-state formation (the BEC limit)
\cite{Nozieres}, which is reflected in the correlation functions.
We now argue that all these phenomena can indeed be observed with
bosonic systems as well. {\bf i)}{\em Gaps for spin and charge
excitations.} The spectrum for the bosonic system is the same that
for Hubbard fermions. Moreover,  spin and charge fermionic
excitations are mapped onto spin and charge bosonic excitations,
since all density (spin) operators $c^{\dagger}_{i \sigma}c_{i
\sigma}$ ($c^{\dagger}_{i \downarrow}c_{i \uparrow}$) are mapped
onto bosonic operators, $a^{\dagger}_{i \sigma}a_{i \sigma}$
($a^{\dagger}_{i \downarrow}a_{i\uparrow}$), by the transformation
(\ref{J-W}). {\bf ii)} {\em Correlation functions.} Since the
fermionic and bosonic amplitudes are related by a sign factor, it
follows that any measurement in our bosonic system involving
densities will give the same result as in the fermionic system,
since the sign factor is then squared. Thus, both density-density
correlation functions $\langle n_{x + \Delta} n_{x} \rangle$,
(where $n_{x}=n_{x \uparrow}+ n_{x \downarrow}$), and  spin-spin
correlation functions $\langle S_{x + \Delta} S_{x} \rangle$,
(where $S_{x}=n_{x \uparrow}- n_{x \downarrow}$), remain unchaged
by (\ref{J-W}).

Note that not all observables are invariant under the
transformation (\ref{J-W}). For example, the sign factor
differentiating bosons and fermions will show up in the one-body
correlation function $\langle a^{\dagger}_{x +\Delta \sigma}a_{x
\sigma}\rangle$, which will be different for fermions and bosons.
We can say that we have a system of bosons that behaves in many
ways as a system of Hubbard fermions, though the bosonic nature of
the real components of the system is not completely hidden and can
be detected with some particular measurements.

The above considerations imply that a variety of phenomena could
be observed for the bosonic system. For example, for $V>0$ one may
use these bosonic systems to observe spin--density separation in
the same way as proposed for fermions \cite{Zoller} by just
exciting locally either the spin or the density with a laser and
looking at the propagation. Alternatively, one can analyze the
excitation spectrum to see the two phononic branches. For $V=0$,
if we include an ``impurity'' atom at some position which strongly
interacts with the rest of the atoms the system will exhibit the
Kondo effect \cite{kondo}.

In view of the current interest in the achievement of the BCS
regime with ultracold atoms, the rest of the letter will be
devoted to the attractive regime $V<0$. Although the problem can
be exactly solved via Bethe Ansatz \cite{Lieb}, a variational
formulation proves to be useful. We take
\begin{equation}
f_{B}(\mathbf{x},\mathbf{y})=\mathrm{sgn}(\mathbf{x},\mathbf{y})
\prod_{i=1}^{N/2}\varphi(x_{i}-y_{i}), \label{bcsb}
\end{equation}
where $\varphi$ is to be determined by minimizing the energy. The
physical meaning of (\ref{bcsb}) can be made more transparent if
we write it in terms of fermionic operators as $ |\Psi_{F} \rangle
= [ \sum_{x,y=1}^{M} \varphi(x-y)\,\, c^{\dagger}_{x \uparrow}
c^{\dagger}_{y \downarrow} ]^{N/2}|0\rangle$ which has been used
in the literature for the fermionic case \cite{Nozieres}, and even
found reasonable agreement with the exact wavefunction in 1D
\cite{Tanaka}. This state describes a condensate of pairs with
spatial wave function $\varphi$, $\Delta=x-y$ being the relative
distance of the pair.

Let us first consider the regime of strong attraction $|V| \gg t$.
Here, the sign factor in (\ref{bcsb}) is the same for all
configurations $(\mathbf{x},\mathbf{y})$. The bosonic paired state
takes then the simple form $| \Psi_{B} \rangle \propto \mathcal{P}
[ \sum_{x=1}^{M} a^{\dagger}_{x \uparrow} a^{\dagger}_{x
\downarrow} ]^{N/2}|0\rangle$,  where $\mathcal{P}$ denotes
projection onto the subspace spanned by (\ref{sub}). In this state
atoms are bound into on-site triplet pairs, and the system can be
visualized as formed of hard-core composite bosons. As it happens
for Hubbard fermions with strong attraction \cite{Nozieres}, the
composite bosons move only via virtual ionization with an
effective hopping amplitude $t_{b} =2t^2/|V|\ll t$. When the
attraction is so strong that we approach the limit $|V|=U$ the
system looses its fermionic character. The state made by on-site
pairs becomes unstable with respect to having more than one pair
on the same site, since the cost in repulsion is exactly
compensated by the gain in attraction. In fact, for $|V|>U$ it is
favorable to have all the particles ($N/2$ with spin $\uparrow$,
$N/2$ with spin $\downarrow$) on the same site. The system behaves
as a single ``big boson'' that moves very slowly along the
lattice. As the ration $|V|/t$ decreases, the function $\varphi$
evolves continuously from a localized state of the Cooper pairs at
$\Delta=0$ to a delocalized one.

\begin{figure}[ht]
\begin{center}
\includegraphics[height=6cm,width=6.5cm]{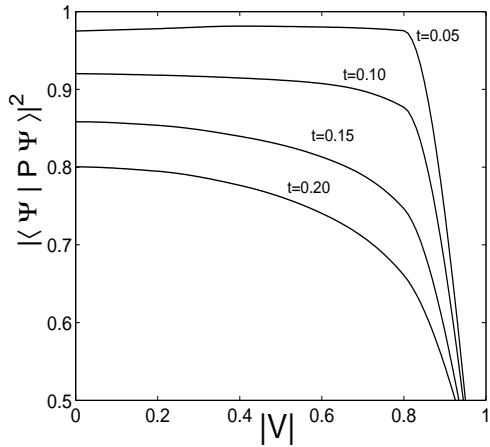}
\end{center}
\caption{Projection of the exact ground state onto the one
corresponding to the limit $U\to \infty$ as a function of $|V|/U$
for $N=6$, $M=8$, and different values of $t/U$.} \label{solapes}
\end{figure}

To illustrate these behaviors we present now numerical exact
results for a system of $N=6$ bosons and $V<0$. Even though this
is a small number, this may closely represent some experimental
situations with atomic systems \cite{Bloch} where the effective
number of atoms in the 1D lattice may be relatively small. Given
that Hamiltonian (\ref{hamb}) is invariant under global spin
rotations, we diagonalize  it within subspaces of fixed total
spin, $S$, and fixed $z$ component of the spin, $S_{z}$. In order
to to check the ``fermionization scheme'' we show in Fig.\
\ref{solapes} the projection of the exact ground state of the
system on the one calculated with the projected Hamiltonian, as a
function of the strength of the attraction, $|V|$. In agreement
with the predictions, when $t \ll U$ and as far as $|V|<U$ the
overlap is more than $95\%$ . However, as we approximate the limit
$|V|=U$, or as $t$ is increased, the system looses the fermionic
character.

\begin{figure}[ht]
\begin{center}
\includegraphics[height=6cm,width=6.5cm]{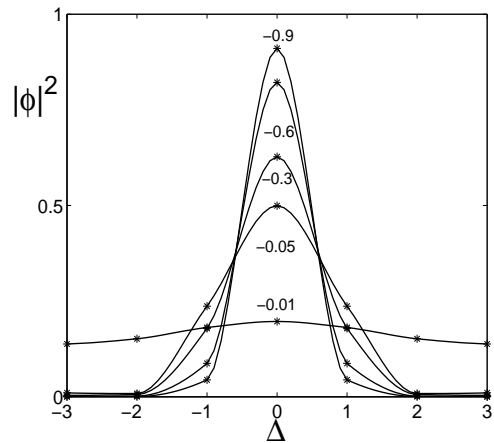}
\end{center}
\caption{$|\varphi|^2$ as a function of $\Delta/U$ for the
variational wave function (\ref{bcsb}) and for $N=6$, $t=0.1U$ and
different values of $|V|/U$.} \label{pairfunc}
\end{figure}

\begin{figure}[ht]
\begin{center}
\includegraphics[height=6cm,width=6.5cm]{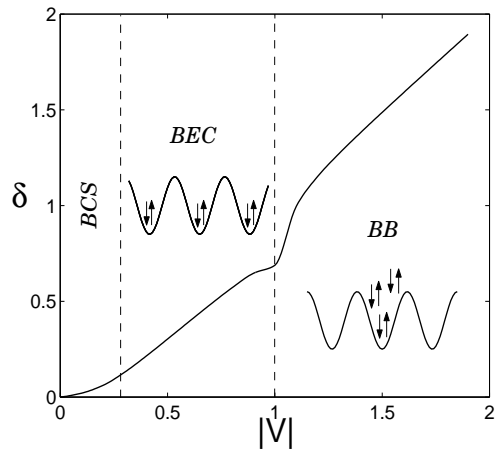}
\caption{Gap for spin excitations, $\delta/U$, as a function of
$|V|/U$ for $N=6$, and $t=0.1U$. The three regimes are indicated
in the figure.} \label{gapt}
\end{center}
\end{figure}

For a regime of parameters in which the fermionization is valid,
we can use the variational wave function (\ref{bcsb}) to
understand the nature of the bosonic pairs. Figure \ref{pairfunc}
shows the probability density of the pair, $|\varphi|^2$,
corresponding to the variational state (\ref{bcsb}). We clearly
see how the nature of the pair exhibits a smooth crossover from
being delocalized, for $|V| \ll t$ to get localize at $\Delta=0$
for $|V| \gg t$. The gap for spin excitations, given by the
expression $\delta=(E_{N-1}-2E_{N}+E_{N+1})/2$ \cite{Lieb}, is
plotted in Fig.\ \ref{gapt} as a function of $|V|$. As the
attraction increases the energy required to break a pair
increases. The behavior of the gap near $|V|=U$ is due to the
transition from a paired state to the ``big boson'' state.

\begin{figure}
\begin{center}
\includegraphics[height=6cm,width=6.5cm]{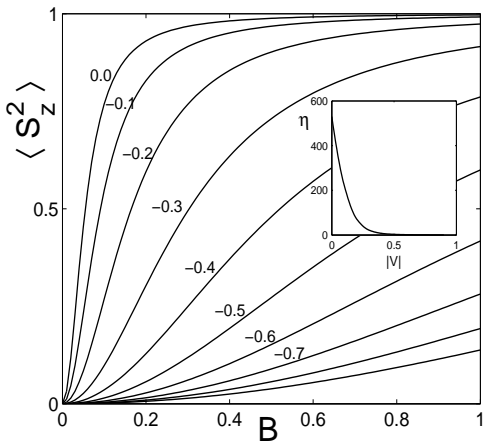}
\caption{$\langle S_{z}^{2} \rangle$ as a function of $B/U$, for
$N=6$, $t=0.1U$, and different values of $V/U$.  The inset shows
the second derivative, $\eta$, of $\langle S_{z}^{2} \rangle$ with
respect to $B$ (in units of $U$) at zero magnetic field, as a
function of $|V|/U$.} \label{medida}
\end{center}
\end{figure}

The existence of pairs and the smooth evolution from Cooper pairs
to localized pairs can be detected by analyzing the response of
the system to a (Raman) laser that couples the internal states of
the atoms, so that $H_{B}=H+ B \sum_{i=1}^{M} a^{\dagger}_{i
\uparrow}\, a_{i \downarrow}+a^{\dagger}_{i \downarrow}\, a_{i
\uparrow}$, where the parameter $B$ can be varied by tuning the
intensity of the laser. The effect of this laser can be understood
as follows. Let us assume that the system is in a paired state
with pair wave function $\varphi(x)(|\uparrow \downarrow \rangle +
|\downarrow \uparrow \rangle)$. The laser will try to rotate the
spin state into the state $|\uparrow \uparrow \rangle +
|\downarrow \downarrow \rangle$. Since this rotation must break
the pair it follows that the response will be very low until a
value $B\sim \delta$ is reached. Thus, $\delta$ can be detected by
observing, for instance, the fluctuations induced in the the $z$
component of the total spin, $\langle S_{z}^{2} \rangle$, as a
function of $B$, as it is illustrated in Fig.\ \ref{medida}. As
predicted, the system response, quantified by $\left. d^2 \langle
S_{z}^{2} \rangle /dB^2 \right|_{B=0}$, is extremely large for
$V=0$, when we have no pairs and the system holds no resistance to
be rotated, and decreases monotonically for increasing attraction,
revealing the existence of pairs. Note that for $|V| \gg t$,
rotation of the on-site pair will imply exciting the system out of
the subspace spanned by (\ref{sub}), since two atoms with the same
spin will be left on the same site, and therefore excitation will
occur for $B\agt U$.

In summary, we have shown that with the experiments that are being
carried out with bosons in optical lattices one can observe a
great variety of fermionic correlation phenomena, including Cooper
pair formation and spin--density separation. These atomic systems
may be simpler to manipulate than fermions themselves and can
provide a very clean laboratory to study several interesting
phenomena which have eluded observation so far. Apart from that,
the system proposed here may bring up novel phenomena in 2 and 3D
optical lattices. Even though the Jordan--Wigner transformation is
not appropriate, some of the phenomena predicted for Fermions
(like d-wave pairing or an analogue) may appear in the case of
bosons in the regime $U> |V|$, and therefore these cases are worth
exploring both theoretically and experimentally.

Discussions with I. Bloch, J. von Delft, C. Tejedor, and P. Zoller
are gratefully acknowledged. B. P. is indebted  to G.
G\'{o}mez-Santos for illuminating discussions.


\begin{thebibliography}{99}
\bibitem{footnotea}
The only requirement is that the interaction constant between the
same internal levels is larger than the one between different
ones. For simplicity we have assumed that both internal levels
have identical interactions, as it would be the case for the
$m=\pm 1$ states of the hyperfine manifold. Note also that in this
regime no three body loses can occur, since there are never three
atoms in the same site.
\end{thebibliography}
\end{document}